\documentclass[conference,10pt]{IEEEtran}
\usepackage[utf8]{inputenc}

\usepackage{graphicx}
\usepackage{amsmath}

\usepackage{todonotes}

\begin{document}

\title{Towards a Fast and Accurate Model of Intercontact Times for Epidemic Routing}

\author{\IEEEauthorblockN{Fabricio Cravo}
\IEEEauthorblockA{\textit{Universit\'e Paris-Saclay, CNRS}\\
\textit{Faculdade de Tecnologia, UNICAMP}\\
fabricio.cravo@student.ecp.fr}
\and
\IEEEauthorblockN{Thomas Nowak}
\IEEEauthorblockA{\textit{Universit\'e Paris-Saclay, CNRS} \\
thomas.nowak@lri.fr}
}

\maketitle

\begin{abstract}
We present an accurate user-encounter trace generator based on analytical models.  Our method generates traces of intercontact times faster than models that explicitly generate mobility traces.  We use this trace generator to study the characteristics of pair-wise intercontact-time distributions and visualize, using simulations, how they combine to form the aggregate intercontact-time distribution. Finally, we apply our trace-generation model to the epidemic routing protocol.  
\end{abstract}


\section{Introduction}

Pocket-switched networks (PSNs) are networks in which mobile agents use small communication devices with a short-range communication protocol.
By using the movement of the members of this network to transport data, we explore the opportunities for connections that arrive from the users' encounters to exchange information. 
In this type of network, paths between nodes are not available at all times.
Therefore, since it is necessary to wait for human encounters to establish connectivity, this network type falls under the category of delay-tolerant networks (DTNs)~\cite{hui2005pocket}.

When studying PSNs, an important parameter is a time between the end of a connection and the start of the next connection~\cite{chaintreau2007impact}. This parameter is called the intercontact time. Several authors have used these parameters to estimate the potential of PSNs since it allows us to estimate the expected time the network must wait for establishing communication between agents. Chaintreau, Hui, Crowcroft, Diot, Gass, and Scott~\cite{chaintreau2007impact} have proposed a power law as a fit for the intercontact time probability distribution and analyzed the performance of naive forwarding protocols under this hypothesis, indicating that we would have an infinite expected time for message arrival. However, Karagiannis, Le Boudec, and Vojnovi\'c~\cite{karagiannis2010power} obtained a finite expected time for message arrival when remarking the exponential decay present in the same traces after the power-law pattern. 

A vital aspect to consider when studying PSN is the heterogeneity of the networks~\cite{passarella2012analysis}. There is a distinction between the aggregate intercontact-time distribution, which is generated by analyzing all intercontact times obtained in an experiment, to the pair-wise intercontact-time distribution, which comes from the intercontact times of single pairs of users. Both Chaintreau et al.~\cite{chaintreau2007impact} and Karagiannis et al.~\cite{karagiannis2010power} assumed that the pair-wise distribution was equal to the aggregate distribution for their analysis.
Even if the aggregate intercontact-time distribution follows a power-law with an exponential decay after the half-day period~\cite{karagiannis2010power}, this does not imply that all pairs follow the same distribution.
More recent research in human mobility patterns tends to indicate that pair-wise distributions are indeed heterogeneous \cite{passarella2011characterising,passarella2012analysis,conan2007characterizing,pajevic2013epidemic} and the aggregate intercontact-time distribution measured in several contact trace experiments is generated by the combination of different distributions. 

Another relevant parameter when dealing with PSNs is the contact rate or the number of contacts per time unit. The contact rate is defined by Hern\'andez-Orallo, Cano, Calafate, and Manzoni~\cite{hernandez2016new} as the limit of the expected number of contacts as a function of time divided by time.
If we model the intercontact times as purely exponential distributions, the expected contact rate becomes the inverse of the scale parameter. However, in this paper, we will approximate this limit for finite times as the number of contacts throughout the trace experiment.

Usually, the evaluation of the performance of opportunistic networks uses computationally expensive mobility models~\cite{hernandez2016new}. Thus we propose in this paper a human encounter trace generator that uses the pair-wise intercontact-time distributions and contact duration probability distributions for the construction of its traces. 

We will show that our proposed model can generate an aggregated intercontact-time distribution similar to the obtained in the real human-mobility experiments. Also, the model can reproduce the contact rates seen in real data. Finally, we were able to introduce preferential meeting times and periodicity to the intercontact times, an aspect of human encounters with is discussed in mobility models~\cite{ekman2008working}.

Moreover, we conducted simulations regarding the behavior of our trace generation in epidemic routing.
By adapting the centrality measurement used by Gao, Li, Zhao, and Cao~\cite{gao2009multicasting}, we did a similar analysis to our traces. The idea behind this is to explore the use of intercontact-time models in digital epidemiology, similar to Pajevic, Karlsson, and Helgason~\cite{pajevic2013epidemic}.
We believe that, by creating traces that can accurately replicate epidemic routing behavior, we could potentially translate those results to modeling disease epidemics. 

For simplicity, we chose not to add a social aspect to intercontact times between a group of distributions and their intercontact-time distributions.
Hui, Chaintreau, Scott, Gass, and Crowcroft~\cite{hui2010bubble} used the social characteristics of communities to improve routing protocols.
Moreover, Yang,  Jiang, et al. \cite{yang2012characterizing} have described a new behavior on intercontact times regarding the alternation between bursts of contacts and extended periods with no contacts, and the importance of weekdays when modeling intercontact times, with weekends having fewer contacts. We leave the incorporation of their findings to this model to future work. 

\textbf{Contribution.}
We make the following contributions in this paper:
\begin{enumerate}
\item We provide a fast and accurate stochastic trace generator for intercontact times in PSNs. For instance, the studied traces on the results were generated under one second on a home computer. However, studies of this application for larger populations have been left to future work.

\item The trace generation provided by this model does not rely on mobility models. Therefore, it does not need to compute users' movement. As far as we know, this is the first work that generates trace experiments from mobility-free analytical models.

\item We compare the model predictions to measured traces from the literature and thereby show its accuracy.

\item We analyze how this model would perform under the epidemic routing protocol as an attempt to link intercontact-time modeling with digital epidemiology. 
\end{enumerate}

\section{Related Work}

Passarela et al. \cite{passarella2011characterising, passarella2012analysis},  studied pair-wise intercontact-time distributions and how they combine to form the observed experimental results on the aggregated distribution. This study also did a hypothesis testing on the contact-rate distributions to analyze how Pareto, gamma, and exponential would perform as potential fits. It has proposed that the exponential distributions we see in mobility data follow a Gamma distribution to create an aggregated power-law.

Narwala et al. \cite{narmawala2014viral} have proposed using community structures and heterogeneous popularity to improve broadcasting. Silvis, Niemeier, and D'Souza. \cite{silvis2006social} analyzed how social ties affect human mobility. Toivonen et al. \cite{toivonen2006model} proposed a model for the construction of social network graphs. Pelusi et al. \cite{pelusi2006opportunistic} provided a good description of the state of the art Opportunistic networks and where this type of network can be useful. Hui et al. \cite{hui2010bubble} created an efficient routing algorithm based on social connections measurements. Wei \cite{wei2013distribution} has analyzed how social interaction impacts the intercontact-time distribution measuring the strength of these social interactions by calculating the total contact duration and k-clique community size.

As examples of experiments that generated intercontact traces, we have the UCSD and Dartmouth data set generated by collecting wireless data from  \cite{mcnett2005access,stonechanging}. The Infocom conference measurements that used contacts from Bluetooth connections in iMotes \cite{hui2005pocket}. The MIT reality mining data set (MIT BT) where a Bluetooth application was able to register logs from other mobile devices \cite{eagle2006reality}. This experiment lasted for 246 days, with 100 devices participating and Bluetooth scans happening every 300 seconds. Therefore, the granularity of this dataset is 300 seconds long. Thus, we cannot record intercontact times and contact durations smaller than this value.

As an example of the mobility model, we have the working day mobility model. This model created routines for humans' daily behaviors and derived their mobility accordingly \cite{ekman2008working}. In this model, they were able to replicate the periodicity and preferential hours of human contacts. 

Pajevic et al. \cite{pajevic2013epidemic} have studied how both homogeneous and heterogeneous analytical intercontact-time models perform under the epidemic routing protocol and discussed the potentials of stochastic intercontact-time modeling in digital epidemiology. Hernandez et al. \cite{hernandez2021human} studied mobility models for Covid epidemic studies.

\section{Description of the Model}

We consider a time-varying graph $G(t) = (V, E(t))$ whose vertices model users carrying mobile communications devices.
At time~$t$, if nodes $v_i$ and $v_j$ are connected, there is an edge between $ v_i $ and $ v_j $.
Let us define the sequence of intercontact times $T^{ij}$ between nodes $v_i$ and $v_j$ as the sequence of random variables $\tau^{ij}_n$.

\begin{equation}
    T^{ij} = \{\tau^{ij}_0, \tau^{ij}_1, \tau^{ij}_2, \tau^{ij}_3, ....\}
\end{equation}

Let us define the contact duration's $T^{ij}_c$ between nodes $v_i$ and $v_j$ as the sequence of random variables $\tau^{ij}_{c,n}$ (as an exception, we need to find the first contact duration as zero, since we assume that no individuals are connected at t equals to 0):

\begin{equation}
    T^{ij}_c = \{\tau^{ij}_{c,0} = 0, \tau^{ij}_{c,1}, \tau^{ij}_{c,2}, \tau^{ij}_{c,3}, ....\}
\end{equation}

We can then define the series of encounter times for the nodes $v_i$ and $v_j$ in the following way:

\begin{equation}
    t^{ij}_{e,n} = \tau^{ij}_n + \tau^{ij}_{c,n} + t^{ij}_{e,(n-1)}
\end{equation}

Now we say that there exists an edge between $ v_i $ and $ v_j $ (the users are connected) at time t when:

\begin{equation}
    t^{ij}_{e,n} < t \leq \tau^{ij}_{c,(n + 1)} + t^{ij}_{e,n} 
\end{equation}

To generate traces, we draw the contact rate for each edge from the inverse of a gamma distribution according to equation~\eqref{eqn:gammadist}.
From this contact rate, we draw the number of expected encounters by multiplying by the simulation duration:

\begin{equation}
\label{eqn:gammadist}
   r_e \sim \mathcal{G}(a, b)
\end{equation}

\begin{equation}
\label{eqn:encounternumber}
    N_e = \lfloor T_{duration} r_e \rfloor
\end{equation}

Using this draw for the number of encounters for each edge the computation of the intercontact times can start.
In this model, two different distributions are proposed to model the pair-wise behavior. By separating the two disjoint groups according to a threshold $T_{e}$ on the contact rate, we assign a power law with exponential decay  \eqref{eqn:powerlawexpdecay} for those with the contact rate above $T_{e}$. To better describe this distribution, it follows a power-law distribution for the intercontact times below a certain threshold $T$ and an exponential distribution afterward. For the edges with a contact rate below $T_{e}$, we assign a uniform distribution. The idea behind this is to represent two types of behaviors, users whose daily routine results in higher contact rates due to matching locations and therefore follow a distribution that allows them to meet more often, and agents who meet sporadically due to unpredicted changes in their routines. The power-law exponential-decay group follows~\eqref{eqn:powerlawexpdecay}, while the other follow~\eqref{eqn:uniform}, as defined below.

If $N_e > T_e$ we first draw a sample $p\sim \mathcal{P}(\alpha)$, then define
\begin{equation}
\label{eqn:powerlawexpdecay}
t_{ic} =
\begin{cases}
p & \text{if } p \leq T\\
D_{day}\mathcal{E} (\lambda) + \mathcal{N}(\mu_{day} - t_{day}, \sigma_{day}) & \text{if } p > T
\end{cases}
\end{equation}
where $\lambda$ and $\alpha$ are chosen so the expected number of encounters for the duration of the experiment matches the assigned number of encounters $N_e$ \cite{passarella2012analysis}. Moreover, $t_{day}$ represents the time of the day in seconds. Here $D_{day}$ is the duration of the day in seconds  $\mathcal{E} (\lambda)$ is the exponential distribution responsible for choosing the number of days they spend apart and $\mathcal{N}(\mu_{day} - t, \sigma_{day})$ is responsible for adding periodicity in seconds to the model allowing agents to meet preferably during day time.

If $N_e \leq T_e$ we draw encounter times from a uniform distribution according to the simulation time. Instead of drawing the intercontact-time samples, we draw the encounter times directly. This is done according to equation~\eqref{eqn:uniform}.

\begin{equation}
\label{eqn:uniform}
   t^{ij}_{e,n}= D_{day}\mathcal{U} (0, D_{sim}) + \mathcal{N}(\mu_{day}, \sigma_{day})
\end{equation}

where $D_{day}$ is the duration of a day in the simulation in seconds, $D_{sim}$ is the duration of days in the simulation in days and $\mathcal{N}(\mu_{day}-t, \sigma_{day})$ is the periodic term.

For the contact durations, that is, the times the edges become active, we simply use a power law according to equation \eqref{eqn:contacttime}

\begin{equation}
\label{eqn:contacttime}
    t_{c} = \mathcal {P}(\alpha_{c}) 
\end{equation}

\begin{figure}
\centering
\includegraphics[scale=0.9]{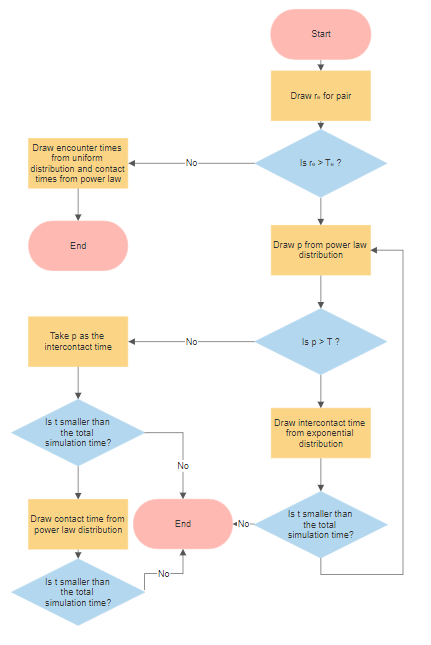}
\caption{Flowchart to describe the pair-wise user-encounter trace generation}
\label{fig:fluxogram}
\end{figure}

Passarela and Conti \cite{passarella2012analysis} have used a Gamma distribution to model contact rates. However, they have only considered pairs of individuals that had more than ten contacts in their analysis. In this work, we use this distribution to model contact rates for all pairs. We do so by fitting the gamma distribution to the number of contacts per pair \cite{conan2007characterizing}. The pair-wise distribution used for pairs above the encounter threshold was a power-law with exponential decay. Because, as indicated by Passarela and Conti \cite{passarella2012analysis},  this distribution has weaker constraints over the contact rate distribution to form the expected aggregate distribution. For the pairs below it, there does not seem to exist literature regarding their analytical analysis. Therefore, we have chosen a uniform distribution to model the unpredictability in their meetings. Finally, we model the contact according to Hui and Chaintreau \cite{hui2005pocket,chaintreau2007impact}. We can find a representation of the model for the pair-wise intercontact-times distributions and contact duration's generations in Figure \ref{fig:fluxogram}.

\section{Results}

\subsection{Model Validation}

\begin{figure}
\centering
\includegraphics[scale=0.5]{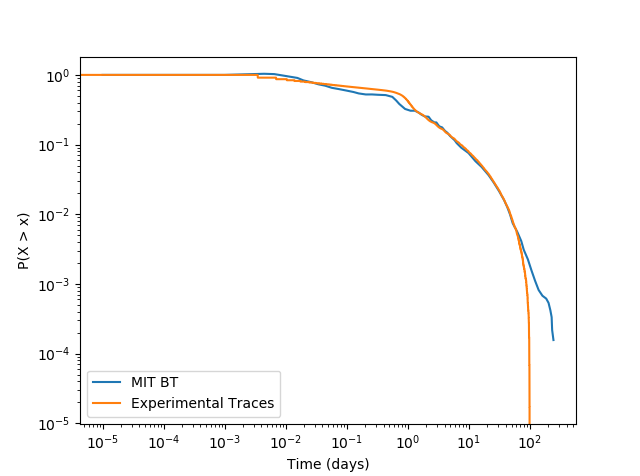}
\caption{Comparison between aggregated intercontact times between MIT BT and model results}
\label{fig:Inter_Contact_Results}
\end{figure}

For the model validation, we compared the generated results with the MIT Reality Mining Bluetooth data \cite{eagle2006reality}. We have used the same amount of agents and used the same time frame as suggested by Conan et al. \cite{conan2007characterizing}  to compare the results. This time frame is the one with the most contacts, which is a period of classes for the students consisting of about 100 days. We generated both the contact duration and intercontact times for each pair according to the model and the inverse CDF of the proposed probability distributions.  However, we only register intercontact times within the time frame of the simulation. Thus, we only consider intercontact times between the end of a contact duration and the start of a new contact duration, to be consistent with datasets measurements. A pair of users had to have at least two contacts to contribute with a single intercontact time for the curve generated in Figure~\ref{fig:Inter_Contact_Results}.
 
\begin{table}
    \centering
        \caption{Model parameters}
 \begin{tabular}{|c c|} 
 \hline
 Parameter & Value \\ 
 \hline\hline
 \#users & 100 \\ 
 \hline
 $D_{sim}$ & 100d \\
 \hline
 $D_{day}$ & 86400s \\
 \hline
 $\mu_{day}$ & 43200s \\
 \hline
 $\sigma_{day}$ & 50 \\
 \hline
 granularity & 300s \\
 \hline
 $T$ & 6030s \\
 \hline
 $a$ & 0.19 \\ 
 \hline
 $b$ & 0.072 \\
 \hline
 $T_{e}$ & 5.79e-7 contacts/s\\
 \hline
\end{tabular}   
    \label{tab:parameters}
\end{table}

The proposed model can replicate well the aggregated intercontact-time distribution from the experimental data.  Firstly, the aggregated distribution generated by the model follows a power-law with an exponential decay distribution with an average relative error of less than 1\% and a maximum relative error of 43\% around the 1-day mark. We can see in both curves the change between the power-law distribution behavior and the exponential distribution behavior happening between the time of half a day and a day. Also, in the MIT BT curve, which is the curve generated by giving volunteers Bluetooth devices that allows one to measure contact duration's and intercontact times \cite{eagle2006reality}, we can see several oscillations on the exponential decay part. These oscillations were replicated by our model due to the addition of $\mathcal{N}(\mu_{day} - t, \sigma_{day})$ which is the part of the model responsible for increasing the likelihood that agents meet during the daytime around the busy hours of the day and reduce the probability of meeting at night.

To guarantee the consistency of contact rates, we compare the average number of encounters per pair generated by the Gamma distribution with the values by  Conan et al. for the MIT BT dataset and obtained an average relative error of 12\% in the number of contacts \cite{conan2007characterizing}. However, we have only compared the pairs of users with at least one contact.  Following the MIT BT dataset, most contact rates generated by the gamma distribution have yielded no contacts throughout the experiment. On average, 51\% of the pairs of users would not encounter under the used parametric conditions.



To study how the trace generator would behave under the pretense of epidemic routing, we generated a set of traces from this model and applied the epidemic routing algorithm. Similar to Pajevic et al. \cite{pajevic2013epidemic}, we consider we start by infecting one random individual, and it infects others as soon as there was a connection available in the traces. Also, we refer to this connection as transmission or infection. Moreover, we say the users are infected once they received a transmission.  The individuals remain infected for the entirety of the broadcast, and we analyze infection rates in the network.  

\begin{figure}
\centering
\includegraphics[scale=0.5]{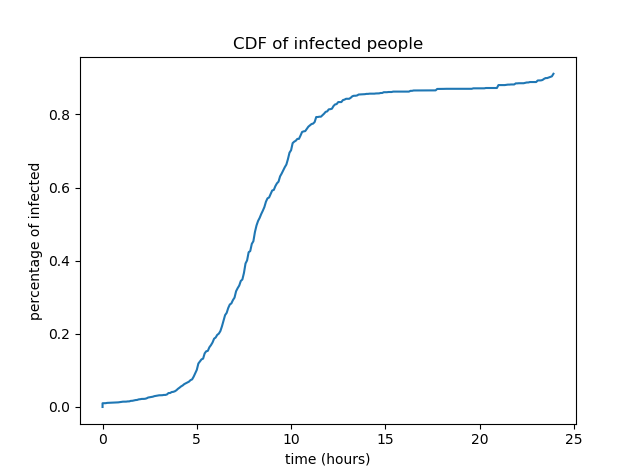}
\caption{Average rates of infection as a function of time}
\label{fig:cdfinfected}
\end{figure}

By choosing a random day and infecting a node at random at 6:00 AM we were able to produce the average infection rates found in Figure~\ref{fig:cdfinfected}. Comparing with the results obtained by Pajevic \cite{pajevic2013epidemic}, it would appear that the broadcasting is extremely slow. However, since the MIT reality mining is the dataset with the lowest amount of contacts per day, it is expected to yield the worst results. Also, Pajevic used the days with the highest amount of contacts for his analysis, while we picked days at random, independently of contact numbers. Using the results recently obtained by Yang \cite{yang2012characterizing}, some weekdays days might have a higher number of contacts due to burst periods. Therefore, using those days for the estimation of epidemic broadcasting would result in an overestimation. 

Gao \cite{gao2009multicasting} has studied how to use social base forwarding to improve routing on multicasting scenarios.  By proposing another metric for measuring centrality, they improve routing efficiency, obtaining results close to epidemic routing. However, they used a smaller number of relays, using fewer network resources than the epidemic case. They modeled the pair-wise number of contacts as a Poisson process, which implies they consider the pair-wise intercontact-time distributions exponential.  Based on their proposed measurement of centrality, we suggest the following measurement, that would be more  adapted to all possible pair-wise distributions:




\begin{equation}
\label{eqn:centralityall}
C_i = 1 - \frac{1}{N - 1} \sum^N_{j = 1, j \neq i} P_{ij}( T > t )
\end{equation}

\begin{figure}
\centering
\includegraphics[scale=0.5]{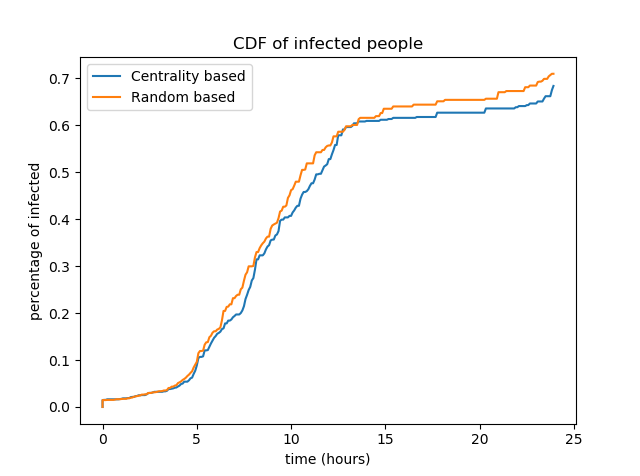}
\caption{Comparison of percentage of infected users by blacklisting at random and centrality based }
\label{fig:cdfblacklist}
\end{figure}

This parameter can be estimated empirically by using measurements of the number of contacts. Or can be calculated using pair-wise intercontact-times distributions. Using equation \eqref{eqn:centralityall}, we decided to blacklist 30 agents with higher centrality values and halt their transmission and compare it with blacklisting 30 agents at random. We start by infecting a node chosen at random and choosing a random day with the starting time of 6:00 am. Afterward, we compare the infection rates for both the random and the centrality-based blacklist and used the average results to generate the distribution in Figure \ref{fig:cdfblacklist}. Moreover, we used the total 100-day data for the computation of the centrality values, even though the broadcasting only happens for one day. Finally,  we have used 6 days for the time $T$ used for centrality calculations.

From these results in Figure \ref{fig:cdfblacklist}, we can see no relevant difference between the two blacklists. If we compare the results obtained by Gao \cite{gao2009multicasting}, and Yoneki \cite{yoneki2008distinct}, we can expect a much more impact on halting the spread of the infection. Indicating the importance of adding a social model to this type of trace generation when modeling epidemic routing. It might be vital to create social hubs and isolated agents as they are of extreme importance when considering broadcasting. By assigning the pair-wise distributions at random we create nodes with highly similar centrality values. 

\begin{figure}[h!]
\centering
\includegraphics[scale=0.5]{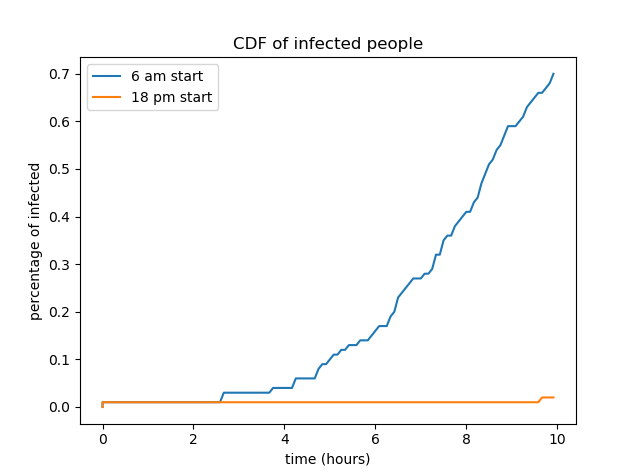}
\caption{Comparison of percentage of infected agents with differing start times}
\label{fig:star_time_comp}
\end{figure}

Finally, we compare broadcasting by using different initial times. We compared the cases where the routing starts at 6:00 to the ones starting at 18:00. The results can be found in Figure \ref{fig:star_time_comp}. Here we can see how the introduced periodicity was able to successfully impact the routing, with the distribution that starts at 18h being well less successful in obtaining higher infection rates.

\subsection{Comparison with Exponential Pair-wise Modeling}

\begin{figure}
\centering
\includegraphics[scale=0.5]{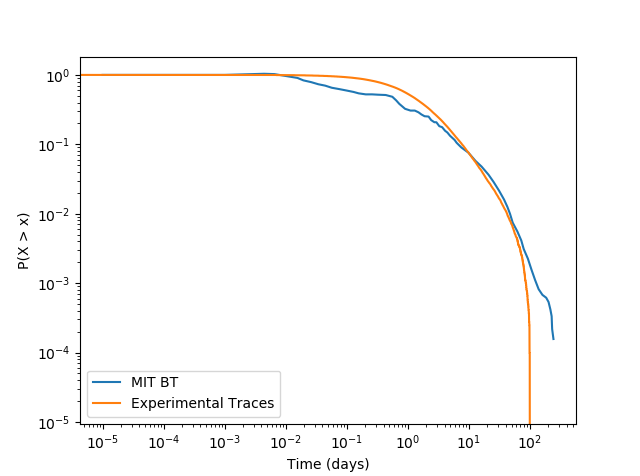}
\caption{Comparison between MIT BT and exponential distribution pair-wise modeling}
\label{fig:exponentialpair-wise}
\end{figure}

Several authors have proposed using exponential distributions for modeling pair-wise intercontact-times distributions \cite{passarella2011characterising, passarella2012analysis, hernandez2016new}. We have also attempted to use this distribution in our model. The idea in this case scenario is that exponential pair-wise distributions whose contact rate is drawn from a gamma distribution form a power-law distribution in the aggregate intercontact-time distribution. Since we already draw the contact from a gamma distribution, it is expected that exponential pair-wise distribution would yield promising results. Therefore, we executed a simulation with the same global parameters for the power-law with exponential decay distribution. However, we change the pair-wise intercontact-time distributions from power-laws with an exponential decay to purely exponential distributions and draw the rate parameter directly from the gamma distribution. The results can be seen in Figure \ref{fig:exponentialpair-wise}. The distribution for the contact rates and parameters for the exponential distributions can be seen in equation \eqref{eqn:gamma}.

\begin{equation}
\label{eqn:gamma}
    p(\lambda) = \frac{\lambda^{\alpha - 1} b^\alpha e^{-b\lambda}}{\Gamma(\alpha)}
\end{equation}

We can see that the exponential pair-wise fit has underperformed the power-law with exponential decay fit. For instance, using the exponential distributions has not yielded enough contacts in the initial power-law region before the one-day mark. Unfortunately, the shapes of the curves did not match. Moreover, we cannot see any indication of the periodicity regarding the contacts. 

\section{Conclusion and Future Work}

In this work, we introduced and analyzed an efficient trace generator for human encounter networks. We also provided evidence for the description of pair-wise intercontact-time distributions as a power-law with an exponential decay after a certain period. Moreover, the evidence seems to indicate that more than one distribution type is essential when considering human interactions. It is expected that this work will aid in the accurate description of pair-wise human intercontact times and allow better simulations for connectivity to be developed and the construction of better models for epidemic routing, aiding in the generation of large-scale intercontact-time simulations. Also, we expect that explicit mobility models will not be used for intercontact-time simulations in the future, because of the expensive computations associated with them.

Unfortunately, our model does not take into account the clustering properties of human social interaction and encounters. Those properties have already been studied by several authors \cite{wei2013distribution,narmawala2014viral,silvis2006social,toivonen2006model,watts1998collective}  and are used in some algorithms to improve the efficiency of routing in PSNs \cite{narmawala2014viral,hui2010bubble}. Considering clustering when generating pair-wise parameters would contribute to the generation of a more accurate model.

It is also necessary to address the problem of the randomness of the pair-wise distributions. Using Yang et al.'s work~\cite{yang2012characterizing} to improve the generation of traces, adding the obtained results for weekend periods and the absence of complete randomness,  could further increase the accuracy of our model.

Finally, this trace generation, using results from analytical models, can prove itself a new way of studying epidemic propagation. Due to the scalability and fast trace generation, it would be possible to apply this type of modeling for larger populations than mobility models. However, a more detailed study that compares classical epidemic literature and intercontact times is required.

\bibliographystyle{plain}
  \fontsize{8.7pt}{9.3pt}\selectfont

\bibliography{bibliography.bib}

\begin{thebibliography}{10}

\bibitem{chaintreau2007impact}
Augustin Chaintreau, Pan Hui, Jon Crowcroft, Christophe Diot, Richard Gass, and
  James Scott.
\newblock Impact of human mobility on opportunistic forwarding algorithms.
\newblock {\em IEEE Transactions on Mobile Computing}, 6(6):606--620, 2007.

\bibitem{conan2007characterizing}
Vania Conan, J{\'e}r{\'e}mie Leguay, and Timur Friedman.
\newblock Characterizing pairwise inter-contact patterns in delay tolerant
  networks.
\newblock In {\em Proceedings of Autonomics 2007}. ACM, New York, 2007.

\bibitem{eagle2006reality}
Nathan Eagle and Alex~Sandy Pentland.
\newblock Reality mining: sensing complex social systems.
\newblock {\em Personal and Ubiquitous Computing}, 10(4):255--268, 2006.

\bibitem{ekman2008working}
Frans Ekman, Ari Ker{\"a}nen, Jouni Karvo, and J{\"o}rg Ott.
\newblock Working day movement model.
\newblock In {\em Proceedings of the 1st ACM SIGMOBILE workshop on Mobility
  models}, pages 33--40, 2008.

\bibitem{gao2009multicasting}
Wei Gao, Qinghua Li, Bo~Zhao, and Guohong Cao.
\newblock Multicasting in delay tolerant networks: a social network
  perspective.
\newblock In {\em Proceedings of the Tenth ACM International Symposium on
  Mobile Ad-Hoc Networking and Computing}, pages 299--308, 2009.

\bibitem{hernandez2021human}
Enrique Hern{\'a}ndez-Orallo and Antonio Armero-Mart{\'\i}nez.
\newblock How human mobility models can help to deal with covid-19.
\newblock {\em Electronics}, 10(1):33, 2021.

\bibitem{hernandez2016new}
Enrique Hern{\'a}ndez-Orallo, Juan~Carlos Cano, Carlos~T Calafate, and Pietro
  Manzoni.
\newblock New approaches for characterizing inter-contact times in
  opportunistic networks.
\newblock {\em Ad Hoc Networks}, 52:160--172, 2016.

\bibitem{hui2005pocket}
Pan Hui, Augustin Chaintreau, James Scott, Richard Gass, Jon Crowcroft, and
  Christophe Diot.
\newblock Pocket switched networks and human mobility in conference
  environments.
\newblock In {\em Proceedings of the 2005 ACM SIGCOMM Workshop on
  Delay-Tolerant Networking}, pages 244--251, 2005.

\bibitem{hui2010bubble}
Pan Hui, Jon Crowcroft, and Eiko Yoneki.
\newblock Bubble rap: Social-based forwarding in delay-tolerant networks.
\newblock {\em IEEE Transactions on Mobile Computing}, 10(11):1576--1589, 2010.

\bibitem{karagiannis2010power}
Thomas Karagiannis, Jean-Yves Le~Boudec, and Milan Vojnovi{\'c}.
\newblock Power law and exponential decay of intercontact times between mobile
  devices.
\newblock {\em IEEE Transactions on Mobile Computing}, 9(10):1377--1390, 2010.

\bibitem{mcnett2005access}
Marvin McNett and Geoffrey~M Voelker.
\newblock Access and mobility of wireless {PDA} users.
\newblock {\em ACM SIGMOBILE Mobile Computing and Communications Review},
  9(2):40--55, 2005.

\bibitem{narmawala2014viral}
Zunnun Narmawala and Sanjay Srivastava.
\newblock Viral spread in mobile social network using network coding.
\newblock In {\em 2014 Annual IEEE India Conference (INDICON)}, pages 1--6.
  IEEE, 2014.

\bibitem{pajevic2013epidemic}
Ljubica Pajevic, Gunnar Karlsson, and {\'O}lafur Helgason.
\newblock Epidemic content distribution: empirical and analytic performance.
\newblock In {\em Proceedings of the 16th ACM International Conference on
  Modeling, Analysis \& Simulation of Wireless and Mobile Systems}, pages
  335--340, 2013.

\bibitem{passarella2011characterising}
Andrea Passarella and Marco Conti.
\newblock Characterising aggregate inter-contact times in heterogeneous
  opportunistic networks.
\newblock In {\em International Conference on Research in Networking}, pages
  301--313. Springer, 2011.

\bibitem{passarella2012analysis}
Andrea Passarella and Marco Conti.
\newblock Analysis of individual pair and aggregate intercontact times in
  heterogeneous opportunistic networks.
\newblock {\em IEEE Transactions on Mobile Computing}, 12(12):2483--2495, 2012.

\bibitem{pelusi2006opportunistic}
Luciana Pelusi, Andrea Passarella, and Marco Conti.
\newblock Opportunistic networking: data forwarding in disconnected mobile ad
  hoc networks.
\newblock {\em IEEE Communications Magazine}, 44(11):134--141, 2006.

\bibitem{silvis2006social}
Julia Silvis, Deb Niemeier, and Raissa D’Souza.
\newblock Social networks and travel behavior: Report from an integrated travel
  diary.
\newblock In {\em 11th International Conference on Travel Behaviour Reserach,
  Kyoto}. Citeseer, 2006.

\bibitem{stonechanging}
Andrew Stone.
\newblock The changing usage of a mature campus-wide wireless network.
\newblock {\em Computer Networks}, 52:2690--2712, 2008.

\bibitem{toivonen2006model}
Riitta Toivonen, Jukka-Pekka Onnela, Jari Saram{\"a}ki, J{\"o}rkki Hyv{\"o}nen,
  and Kimmo Kaski.
\newblock A model for social networks.
\newblock {\em Physica A: Statistical Mechanics and its Applications},
  371(2):851--860, 2006.

\bibitem{watts1998collective}
Duncan~J Watts and Steven~H Strogatz.
\newblock Collective dynamics of ‘small-world’networks.
\newblock {\em Nature}, 393(6684):440--442, 1998.

\bibitem{wei2013distribution}
Kaimin Wei, Renyong Duan, Guangzhou Shi, and Ke~Xu.
\newblock Distribution of inter-contact time: An analysis-based on social
  relationships.
\newblock {\em Journal of Communications and Networks}, 15(5):504--513, 2013.

\bibitem{yang2012characterizing}
Lintao Yang, Hao Jiang, Sai Wang, Lin Wang, and Yuan Fang.
\newblock Characterizing pairwise contact patterns in human contact networks.
\newblock {\em Ad Hoc Networks}, 10(3):524--535, 2012.

\bibitem{yoneki2008distinct}
Eiko Yoneki, Pan Hui, and Jon Crowcroft.
\newblock Distinct types of hubs in human dynamic networks.
\newblock In {\em Proceedings of the 1st Workshop on Social Network Systems},
  pages 7--12, 2008.

\end{thebibliography}

\end{document}